\documentclass[12pt,reqno,a4paper]{article}
\usepackage{amsmath,amsfonts,amssymb,graphicx,epsfig,pict2e,rotating}
\usepackage[english]{babel}
\usepackage[noxcolor]{pstricks}
\usepackage{hyperref}
\numberwithin{equation}{section}

\newcommand{\be}{\begin{equation}}
\newcommand{\ee}{\end{equation}}
\newcommand{\bea}{\begin{eqnarray}}
\newcommand{\eea}{\end{eqnarray}}

\newcommand{\uno}{{\mathbf 1}}

\newcommand{\e}{{\mathbf e}}

\begin{document}
\setcounter{page}{1}

\vspace{8mm}
\begin{center}
{\Large {\bf A q-Virasoro Algebra at roots of unity, Free Fermions and Temperley Lieb Hamiltonians  }}

\vspace{10mm}
 {\Large Alessandro Nigro\footnote{Email: Alessandro.Nigro@mi.infn.it}\\
 Dipartimento di Fisica and INFN- Sezione di Milano\\
 Universit\`a degli Studi di Milano I\\
 Via Celoria 16, I-20133 Milano, Italy}\\
 [.4cm]

  \end{center}

\vspace{8mm}
\centerline{{\bf{Abstract}}}
\vskip.4cm
\noindent
In this work we introduce a novel q-deformation of the Virasoro algebra expressed in terms of free fermions, we then realize that this algebra, when the deformation parameter is a root of unity can be realized exactly on the lattice. We then study the relations existing between this lattice deformed Virasoro algebra at roots of unity and the tower of commuting Temperley-Lieb hamiltonians introduced in a previous work.
\renewcommand{\thefootnote}{\arabic{footnote}}
\setcounter{footnote}{0}

\section{Introduction}
In 2 previous papers on the Ising model by the author \cite{nigro}\cite{nigroTL}, the integrals of motion (IOM) of Bhazanov Lukyanov and Zamoldchikov (BLZ) \cite{blz} were analized by means of Themodynamic Bethe ansatz, and in the most recent work the model on a chain with spatially periodic boundary conditions was analyzed by introducing a finitized version of the BLZ integrals of motion, taking values in the enveloping algebra of the periodic Temperley Lieb algebra with periodic boundaries. In this work it was recognized that the eigenvalues for this finitized IOM admit expansions in $1/N$ (where $N$ is the size of the system) whose coefficients are related to the continuum eigenvalues of the BLZ IOM. This property is well known to be a general property of deformed algebras such as  the Deformed Virasoro Algebra (DVA) \cite{dva}, when one expands the generators around the CFT point in powers of a suitable $\hbar$ which in our case is recognized to be a multiple of the inverse size of the system.\\
These deformed algebra admits a free fermion point, where its representation can be described in terms of deformed fermionic oscillators, this is the case of the DVA when the deformation parameter \cite{dva} takes a special value.\\
The idea of this work started by the analysis of the free fermion point of \cite{dva}, which is realized as a deformed free fermion. It was originally expected that the expansion of the eigenvalues of the finitized IOM would be related to that obtained from the fermionic DVA current of \cite{dva}. However, it became clear that, although the expression for the DVA current  describing the expansions had to be identical, the fermi mode anticommutation relations did not need to be deformed. This led to a different kind of deformed Virasoro Algebra, whose commutations relations the author has not been able to track back to known algebras . The situation seems however not to be new, there is a deformed algebra that acts on undeformed modules over the standard Virasoro algebra. 
It will also be discussed how the Virasoro Algebra emerges from the limit in which the deformation parameter goes to 1.
Finally we will introduce a lattice realization of this deformed Virasoro at roots of unity which will be used to give expressions of the finitized IOM in terms of the modes of a free fermion field, and the deformation parameter will be taken to be a root of unity, the order of this root being related to the size  of the system.\\

\section{Generalities}
In this section we remind some basic facts about the Virasoro algebra at $c=1/2$ and free fermions, which can be found for example in \cite{ginsparg}, in the next section we mimic this construction for a generating field which is different from the stress energy tensor, thus leading to a deformed Virasoro algebra.\\
We consider a free Fermi field in 2 euclidean space-time dimensions
\be \psi(z)=\sum_{n\in\mathbb{Z}-\frac{\delta}{2}}\frac{\psi_n}{z^{n+\frac{1}{2}}}  \ee
where the fermi modes satisfy anticommutation relations
\be  \{\psi_n,\psi_m\}=\delta_{n+m,0}  \ee
to this field is associated a spin 2 stress energy tensor
\be T(z)=-\frac{1}{2}:\psi(z)\partial\psi(z):  \ee
when this field is expanded in modes
\be T(z)=\sum_{n\in\mathbb{Z}}\frac{L_n}{z^{n+2}}   \ee
it is well known that the modes satisfy a Virasoro algebra
\be  [L_n,L_m]=(n-m)L_{n+m}+\frac{c}{12}n(n^2-1)\delta_{n+m,0}   \ee
with central charge $ c=\frac{1}{2}$.
the Virasoro generators admit simplified expressions.\\
For $\delta=0$
\be  L_{2k}=-\frac{1}{2}\sum_{s=1}^\infty (2s)\psi_{k-s}\psi_{k+s} \ee
\be  L_{2k+1}=-\frac{1}{2}\sum_{s=1}^\infty (2s-1)\psi_{k-s+1}\psi_{k+s} \ee
whereas for $\delta=1$
\be  L_{2k}=-\frac{1}{2}\sum_{s=\frac{1}{2}}^\infty (2s)\psi_{k-s}\psi_{k+s} \ee
\be  L_{2k+1}=-\frac{1}{2}\sum_{s=\frac{1}{2}}^\infty (2s+1)\psi_{k-s}\psi_{k+s+1} \ee
The representations of the above Virasoro algebra can be obtained by acting with linear combinations of products of Virasoro modes on suitable highest weight states, in the case of $\delta=1$ the highest weight state is the vacuum state $\big|0\big>$ of conformal weight $h=0$, whereas for $\delta=0$ we have 2 highest weight states $\big|1/16\big>_\pm$ of conformal weight $h=1/16$ which are connected by the action of the zeromode algebra:
\be \psi_0^2=\frac{1}{2}\ee
\be \psi_0\big|\frac{1}{16}\big>_\pm=\frac{1}{\sqrt{2}}\big|\frac{1}{16}\big>_\mp \ee

\section{The q-Virasoro Current}
Motivated by the free fermion point analysis of \cite{dva}, we introduce the following conserved spin 1 current:
\be D(z)= :\psi(q^{-1}z)\psi(qz):   \ee
which can be expanded in modes as:
\be  D(z)=\sum_{l\in\mathbb{Z}}\frac{D_l(q)}{z^{l+1}}   \ee
where the modes are defined as:
\be  D_l(q)= q^{l}\sum_{m\in\mathbb{Z}-\frac{\delta}{2}}q^{-2m}:\psi_{l-m}\psi_{m}:  \ee
notice however that differently from \cite{dva} the fermi modes anticommutation relations are not deformed.\\
The modes of the deformed current can be simplified as:
for $\delta=0$
\be  D_{2k}(q)=(q^{-1}-q)\sum_{s=1}^\infty [2s]_{q}\psi_{k-s}\psi_{k+s} \ee
\be  D_{2k+1}(q)=(q^{-1}-q)\sum_{s=1}^\infty [2s-1]_{q}\psi_{k+1-s}\psi_{k+s} \ee
for $\delta=1$
\be  D_{2k}(q)=(q^{-1}-q)\sum_{s=\frac{1}{2}}^\infty [2s]_{q}\psi_{k-s}\psi_{k+s} \ee
\be  D_{2k+1}(q)=(q^{-1}-q)\sum_{s=\frac{1}{2}}^\infty [2s+1]_{q}\psi_{k-s}\psi_{k+s+1} \ee
where 
\be [n]_q= \frac{q^{-n}-q^{n}}{q^{-1}-q}   \ee
we notice that for $q\to 1$ 
\be [n]_q\sim n+\frac{1}{6}n(n^2-1)\log^2(q)+o(\log^2(q))  \ee
notice that in the limit $q\to 1$ one has
\be D_n\sim 2\log(q^2)L_n+\ldots  \ee
we now want to compute the commutation relations for the deformed modes. For this purpose we compute the operator product expansion between the $D$ currents by means of the Wick theorem for fermionic fields:
\be D(z)D(w)\sim -\frac{q D^+(w)}{z-q^2w} -\frac{q^{-1}D^-(w)}{z-q^{-2}w}+\frac{1}{(z-q^{-2}w)(z-q^2w)}-\frac{1}{(z-w)^2}+\ldots   \ee
where the dots stand for terms that are always regular as $z$ approaches any of the OPE singularities, and $D^{\pm}(w)$ are defined as:
\be D^{\pm}(w)=:\psi(q^{\pm3}w)\psi(q^{\mp1}w):   \ee
which can be expanded on the deformed virasoro modes as:
\be D^{\pm}(w)=\sum_{m\in\mathbb{Z}}\frac{D^{\pm}_m}{w^{m+1}}  \ee
where
\be  D^{\pm}_m= q^{\mp(m+1)}D_m(q^{\mp2})   \ee
we then introduce the quantity:
\be A_{n,m}=\oint\frac{dw}{2\pi i}w^m\oint_{C_w}\frac{dz}{2\pi i}z^n D(z)D(w)  \ee
where the integration contour $C_w$ encircles all the 3 singularities for $z\to w,q^2w,q^{-2}w$. It is then a standard calculation of contour integrals to show that:
\be  A_{n,m}=(q-q^{-1})[n-m]_{q}D_{n+m}(q^2)+([n]_{q^2}-n)\delta_{n+m,0}   \ee
we notice that the $A_{n,m}$ are antisymmetric in $n,m$, therefore the commutation relations of the $D_n$ are then given by:
\be  [D_n(q),D_m(q)]=A_{n,m}  \ee
namely
\be  [D_n(q),D_m(q)]=(q-q^{-1})[n-m]_{q}D_{n+m}(q^2)+([n]_{q^2}-n)\delta_{n+m,0} \ee
we notice that performing an expansion as $q\to1$ we get to lowest nontrivial order
\be  4\log^2(q^2)[L_n,L_m]=\log(q^2)(n-m)2\log(q^4)L_{n+m}+\frac{1}{6}n(n^2-1)\log^2(q^2)\delta_{n+m,0} \ee
which are simplified precisely to the Virasoro commutation relations for $c=1/2$:
\be  [L_n,L_m]=(n-m)L_{n+m}+\frac{1}{24}n(n^2-1)\delta_{n+m,0} \ee
It is also possible to be more general and compute the commutation relations for different $q$ of the modes, one starts from the OPE:
\be \begin{split} D(z,q)D(w,t)&\sim -\frac{tD(tz,tq^{-1})}{w-tq^{-1}z}+\frac{t^{-1}D(t^{-1}z,(tq)^{-1})}{w-(tq)^{-1}z}+\frac{tD(tz,tq)}{w-tqz}-\frac{t^{-1}D(t^{-1}z,qt^{-1})}{w-qt^{-1}z}+\\&-\frac{1}{(w-tq^{-1})(w-qt^{-1}z)}+\frac{1}{(w-(tq)^{-1}z)(q-tqz)} \end{split}\ee
and as done previously computes
\be [D_n(q^\alpha),D_m(q^\beta)]=\oint\frac{dz}{2\pi i}z^n\oint_{C_w}\frac{dw}{2\pi i}w^m D(z,q^\alpha)D(w,q^\beta)  \ee
to obtain the commutation relations
\be\begin{split} [D_n(q^\alpha),D_m(q^\beta)]&=(q-q^{-1})[\alpha m-\beta n]_q D_{m+n}(q^{\alpha+\beta})-(q-q^{-1})[\alpha m+\beta n]_q D_{m+n}(q^{\alpha-\beta})+\\&+([m]_{q^{\alpha+\beta}}-[m]_{q^{\alpha-\beta}})\delta_{n+m,0} \end{split} \ee
finally we want to obtain the commutation relations between the q-Virasoro modes and the free fermion modes, in order for example to be able to study null vectors in the representation space and generic eigenstates of the deformed dilatation generator in their q-Virasoro form.\\
We start by considering the operator product expansion of the DVA current with the free fermion  field:
\be D(z,q)\psi(w)=\frac{\psi(qz)}{q^{-1}z-w}-\frac{\psi(q^{-1}z)}{qz-w} \ee
it follows that the commutator between the $D_n$ and the fermi field is:
\be [D_n,\psi(w)]=\oint\frac{dz}{2\pi i}z^n D(z)\psi(w)  \ee
where the integration contour encloses both the singularities of the OPE, so that one has
\be [D_n,\psi(w)]=(qw)^{n} q\psi(q^2w)-(q^{-1}w)^n q^{-1}\psi(q^{-2}w)  \ee
from which it is easy to find that:
\be [D_n,\psi_m]=(q^{-1}-q)[n+2m]_q\psi_{m+n}  \ee
as an example we compute the fermionic form of the generic level 2 descendant of the state with conformal weight $1/16$:
\be (a D_{-2}+b D_{-1}^2)\big|\frac{1}{16}\big> =(a (q^{-1}-q)[2]_q+b(q^{-1}-q)^2[-3]_q)\psi_{-2}\psi_0\big|\frac{1}{16}\big>      \ee
from which it follows that we have the following null state at level 2
\be ((q^{-1}-q)[3]_q D_{-2}+[2]_q D_{-1}^2)\big|\frac{1}{16}\big> =0   \ee
in general it is possible to use the same method for computing all the null vectors, and in view of the next section all the eigenstates of the deformed dilatation generator in terms of the deformed modes $D_n$. We shall see that this method can be applied straightforwardly on the lattice as well.\\ 

   \section{q-Virasoro and CFT Integrals of Motion}
We define the following normal ordered quantities
\be  I_{m}=\sum_{n\in\mathbb{Z}-\frac{\delta}{2}}n^{m}:\psi_{-n}\psi_n:   \ee
by straightforward manipulation of the normal ordering one sees that
\be I_{2k}=0   \ee
whereas:
\be\label{IOM}  I_{2k-1}=2\sum_{n=1-\frac{\delta}{2}}^\infty n^{2k-1}\psi_{-n}\psi_n  \ee
The above quantities coincide up to a suitable c-numeric shift and rescaling by a constant with the local integrals of motion of \cite{blz}, therefore by abuse of language we shall call them local integrals of motion as well. Let us now consider the problem of writing down their eigenstates and eigenvalues.\\
If we consider a state characterized by a fermionic partition $\mathcal{P}$, in the sector where $\delta=1$ one has:
\be \big|\mathcal{P}\big>=\prod_{k\in\mathcal{P}}\psi_{-k+\frac{1}{2}}\big|0\big>   \ee
while for $\delta=0$ we have
\be \big|\mathcal{P}\big>=\prod_{k\in\mathcal{P}}\psi_{-k}\big|\frac{1}{16}\big>_\pm   \ee
it is then straightforward to see that the eigenvalues of the integrals of motion are given by:
\be  I_{2k-1}\big|\mathcal{P}\big>=2\sum_{n\in\mathcal{P}} \bigg(n-\frac{\delta}{2}\bigg)^{2k-1}\big|\mathcal{P}\big>  \ee
We remark that the eigenstates of the IOM can be easily expressed in terms of Virasoro modes,  for example the sixth level of descendance in the vacuum sector we have:
\be 20L_{-6}\big|0\big>+24L_{-4}L_{-2}\big|0\big>+5L_{-3}^2\big|0\big>=112\psi_{-\frac{11}{2}}\psi_{-\frac{1}{2}}\big|0\big> \ee
\be 4L_{-6}\big|0\big>-8L_{-4}L_{-2}\big|0\big>+L_{-3}^2\big|0\big>=16\psi_{-\frac{9}{2}}\psi_{-\frac{3}{2}}\big|0\big> \ee
\be 20L_{-6}\big|0\big>+24L_{-4}L_{-2}\big|0\big>-23L_{-3}^2\big|0\big>=56\psi_{-\frac{7}{2}}\psi_{-\frac{5}{2}}\big|0\big> \ee
We now consider the q-Virasoro $D_0$ generator, it is then straightforward to see that it can be expanded as:
\be  D_0=\sum_{k=1}^\infty \frac{\log^{2k-1}(q^2)}{(2k-1)!}I_{2k-1}  \ee
we now notice that if we set
\be q^2=e^{i\frac{\pi}{D}}  \ee
the expansion for $D_0$ closely resembles the expansion of the eigenvalues TL hamiltonian ${\bf A}_1$ of \cite{nigroTL} in inverse powers of the size $D$ up to an overall c-numeric shift ${\bf A}_1\to {\bf A}_1-c \uno$. This suggests the idea of introducing a lattice realization of the q-Virasoro algebra at roots of unity, which will be carried out in the next section. 

\section{q-Virasoro at roots of unity on the lattice }
In this section we are going to introduce the q-Virasoro algebra at roots of unity directly on the lattice and to express the lattice involutive hamiltonians, introduced in \cite{nigroTL}, directly in terms of the deformed dilatation generator.\\
We first introduce the Clifford algebra of 2D-dimensional euclidean space:
\be   \{\Gamma_i,\Gamma_j\}=\delta_{i,j}  \ee
where the $\Gamma$s have dimension $2^D\times 2^D$. And use it to build a representation of the Temperley-Lieb algebra \cite{saleur}:
\be  \e_j=  \frac{1}{\sqrt{2}}+i \sqrt{2}\Gamma_j \Gamma_{j+1}     \ee
which satisfies the relations:
\be \e_i^2=\sqrt{2}\e_i  \ee
\be  \e_i\e_{i\pm 1}\e_i=\e_i      \ee
\be \e_i\e_j=\e_j\e_i \ , \ |i-j|\geq 2  \ee
where we require periodic boundary conditions $\Gamma_{i+2D}=\Gamma_i$, we then introduce the Fourier transform of the gamma matrices, which provides a lattice discretization of the continuum Fermi modes $\psi_n$:
 \be   \psi_k=\frac{1}{\sqrt{2D}}\sum_{j=1}^{2D}\Gamma_j e^{j k \frac{i\pi}{D}}           \ee
satisfying
\be  \{\psi_n,\psi_m\}=\delta_{n+m,0}  \ee
and the following periodicity
\be \psi_{n+2D}=\psi_n   \ee
now take
\be   q=e^{i\tau}   \ee
\be \tau= \frac{\pi}{2D}  \ee
and introduce the following quantities:
\be  D_{2k}(q)=(q^{-1}-q)\sum_{s=1}^{2D} [2s]_{q}\psi_{k-s}\psi_{k+s}-\frac{1}{q-q^{-1}}\delta_{2k,0} \ee
\be  D_{2k+1}(q)=(q^{-1}-q)\sum_{s=1}^{2D} [2s-1]_{q}\psi_{k+1-s}\psi_{k+s}-\frac{1}{q-q^{-1}}\delta_{2k+1,0}  \ee
which satisfy
\be\begin{split} [D_n(q^\alpha),D_m(q^\beta)]&=(q-q^{-1})[\alpha m-\beta n]_q D_{m+n}(q^{\alpha+\beta})-(q-q^{-1})[\alpha m+\beta n]_q D_{m+n}(q^{\alpha-\beta})+\\&+([m]_{q^{\alpha+\beta}}-[m]_{q^{\alpha-\beta}})\delta_{n+m,0} \end{split} \ee
we now consider the following Temperley-Lieb hamiltonians \cite{nigroTL}:
\be  \mathbf{A}_1={\bf H}_1-2D\uno \ee
\be \mathbf{A}_3={\bf H}_{3}+6\mathbf{A}_1  \ee
\be   \mathbf{A}_5=6{\bf H}_{5}+60\mathbf{A}_3-120\mathbf{A}_1  \ee
\be  \mathbf{A}_7=90{\bf H}_{7}+210\mathbf{A}_5-5040\mathbf{A}_3+5040\mathbf{A}_1   \ee
\be  \mathbf{A}_9= 2520{\bf H}_{9}+504\mathbf{A}_7-45360\mathbf{A}_5+604800\mathbf{A}_3-362880\mathbf{A}_1    \ee
\be \mathbf{A}_{11}=113400{\bf H}_{11}+990\mathbf{A}_9-221760\mathbf{A}_7+11642400\mathbf{A}_5-99792000\mathbf{A}_3+39916800\mathbf{A}_1  \ee
and in general:
\be \mathbf{A}_{2k-1}=\frac{(2k-2)!}{2^{k-1}}{\bf H}_{2k-1}+\sum_{m=1}^{k-1}\frac{(-1)^{m+1}}{m}\binom{2k-m-2}{m-1}\frac{(2k-1)!}{(2k-2m-2)!}\mathbf{A}_{2m-1}  \ee
being
\be  {\bf H}_{2k-1}=\sqrt{2}\sum_{n=1}^{2D}[\e_n,[\e_{n+1},[\e_{n+2},\ldots[\e_{n+2k-3},\e_{n+2k-2}]\ldots]]]   \ee
the relevant result is that the above TL hamiltonians can be decomposed in terms of the $D_0$ generator at roots of unity in the following form:
\be  \mathbf{A}_{2n-1}=\frac{i (4n-4)!!}{2^{2n-3}}\sum_{k=0}^{n-1}\binom{2n-1}{k}(-1)^{n-k-1}\Big(D_0(q^{2(n-k)-1})+\frac{1}{q^{2(n-k)-1}-q^{-(2(n-k)-1)}}\Big)     \ee
another expression directly in terms of the free fermi modes reads:
\be    \mathbf{A}_{2n-1}=2(4n-4)!!\sum_{k=1}^{2D}\sin^{2n-1}(\frac{\pi k}{D})\psi_{-k}\psi_{k}  \ee
whereas the eigenvalues of the lattice IOM are found to be given by the following formula:
\be    A_{2n-1}=2(4n-4)!!\Big(2\sum_{k\in \mathcal{P}}\sin^{2n-1}(\frac{\pi k}{D}) -\sum_{k=1}^{D}\sin^{2n-1}(\frac{\pi k}{D}) \Big) \ee
where the fermionic partition $\mathcal{P}$ is restricted in the sense that each of its elements must be less or equal than $D$, and it is said to be fermionic because its entries do not repeat themselves, otherwise in the corresponding state a fermi mode would repeat itself twice and thus would give zero . The eigenstates of the lattice IOM are then of the following form:
\be \big|\mathcal{P}\big>=\prod_{k\in\mathcal{P}}\psi_{-k}\sigma_\pm  \ee
where $\sigma_\pm$ is a state such that
\be \psi_n\sigma_\pm=0 , \quad  0<n<D  \ee
the states $\sigma_\pm$ can be thought as a lattice analogue of the highest weight state $\big|\frac{1}{16}\big>_\pm$. Obviously the two states are connected by the action of the lattice zeromode algebra:
\be \psi_0^2=\frac{1}{2}\ee
\be \psi_0\sigma_\pm=\frac{1}{\sqrt{2}}\sigma_\mp  \ee
It turns out that one can also build the eigenstates and the null states on the lattice in terms of the  lattice $D_n$ at roots of unity by using the lattice commutation relations
\be [D_n,\psi_m]=(q^{-1}-q)[n+2m]_q\psi_{m+n}  \ee
 so that for example we have the following null state which is completely analogous to the continuum counterpart:
\be ((q^{-1}-q)[3]_q D_{-2}+[2]_q D_{-1}^2)\sigma_\pm=0   \ee

\section{Conclusions}
In this paper we have introduced a q-deformation of the Virasoro algebra, we have then checked that in the limit $q\to 1$ we get to the first nontrivial order of the expansion the Virasoro algebra.
We have then realized that this algebra, when the deformation parameter is a root of unity can be exactly realized on the lattice in terms of the clifford algebra of $\Gamma$ matrices. This lattice realization enjoys several nice properties, such as the existence of exact null vectors on the lattice, and especially it is related to the Temperley-Lieb hamiltonians of \cite{nigroTL} which can be decomposed in a simple way in terms of the deformed dilatation generator at roots of unity. The eigenvalues of the Temperley-Lieb hamiltonians are found, and their simple form can be easily guessed from their simple form in terms of lattice fermi modes.\\
The TL hamiltonians turn out to be given by the same formulas as in \cite{nigroTL} although the content in terms of representations of the Virasoro algebra is different, this provides an important consistency check for the conjectural closed form relation of the lattice IOM obtained in \cite{nigroTL} which should be actually independent from the decomposition in terms of representations.\\

\end{document}